\documentclass[showpacs,amsmath,twocolumn,showkeys,pra]{revtex4-1}
\usepackage{dcolumn}    
\usepackage{bm}         
\usepackage{ifpdf}
\usepackage{amssymb,lineno,amsfonts}
\usepackage{graphicx}   
\usepackage{bbm}
\usepackage{mathrsfs}
\usepackage{upgreek}
\usepackage{mathtools}
\usepackage{epstopdf}
\usepackage{setspace}
\usepackage{hyperref}
\usepackage[matrix,frame,arrow]{xypic}
\usepackage{rotating}
\usepackage{float}
\usepackage{natbib}
\usepackage{color} 
\definecolor{med-blue}{RGB}{25,25,112} 
\hypersetup{colorlinks, linkcolor={red},citecolor={blue}, urlcolor={blue}}

\newcommand{\ket}[1]{\vert{#1}\rangle}

\newcommand{\outpr}[2]{\vert{#1}\rangle\langle{#2}\vert}

\newcommand{\expv}[1]{\langle{#1}\rangle}
\newcommand{\mprod}[3]{\langle{#1}\vert{#2}\vert{#3}\rangle}

\begin{document}
\title{Estimating Franck-Condon factors using an NMR quantum processor}
\author{$^1$Sharad Joshi, $^1$Abhishek Shukla, Hemant Katiyar, $^2$Anirban Hazra, and $^1$T. S. Mahesh}
\email{mahesh.ts@iiserpune.ac.in}
\affiliation{$^1$Department of Physics and NMR Research Center,\\
$^2$Department of Chemistry,\\
Indian Institute of Science Education and Research, Pune 411008, India}

\begin{abstract}
{
Interaction of molecules with light may lead to 
electronic transitions and simultaneous vibrational excitations.
Franck-Condon factors (FCFs) play an important role in quantifying the intensities of such 
vibronic transitions occurring during molecular photo-excitations.
In this article, we describe a general method for estimating FCFs
using a quantum information processor.  The method involves the application of a translation
operator followed by the measurement of certain projections. 
We also illustrate the method
by experimentally estimating FCFs with the help of a three-qubit
NMR quantum information processor. We describe two methods for the
measurement of projections - (i) using diagonal tomography and 
(ii) using Moussa protocol.  The experimental results agree fairly well with
the theory.
}\end{abstract}

\maketitle

\section{Introduction}

Electronic transitions in molecules are often associated with
transitions in vibrational levels, and such combined transitions are known
as vibronic transitions \cite{herzberg1950spectra}.  
In a vibronic transition, the displacements of the nuclei, owing to their higher masses,
occur at a much slower pace than the electronic rearrangements. 
This phenomenon is contained in the Franck-Condon principle, which states that
the transition probability between two vibronic levels depends on the 
overlap between the respective vibrational wavefunctions \cite{demtroder2006atoms}.  
Thus, in a vibronic transition, the intensities of the absorption or emission spectra are dictated by Franck-Condon factors (FCFs) which are the squared magnitudes of the overlaps of vibrational wavefunctions. 
Estimating FCFs is an important task in understanding 
absorption and fluorescence spectra and related phenomena such as photo-induced dissociations \cite{bowers1984fragmentation}.  

In this work, we estimate FCFs using a three-qubit Nuclear Magnetic Resonance (NMR) 
quantum processor \cite{cory2000nmr,suter2008spins}.  Long coherence times and precise control on spin-dynamics 
make NMR systems ideal testbeds for studying various aspects of quantum information
and quantum physics \cite{oliveira2011nmr}.  For example, probability distributions of a particle
in various potentials has recently been studied using a five-qubit NMR processor \cite{shankar2014quantum}.
We describe an elegant method based on the iterative 
application of a translation operator to achieve an arbitrarily high spatial resolution 
of probability distributions. 

Here we estimate FCFs 
corresponding to a pair of harmonic oscillators as a function of
their relative displacements.
An important step in this method is the measurement of certain projection
operators.  
In this article we illustrate two different approaches for the measurement, namely
(i) diagonal density matrix tomography and (ii) Moussa protocol.

In the following section we briefly describe the theory of FCFs and their estimation.
The experimental results are described in section III and we conclude in section IV.

\section{Theory}
\subsection{Franck-Condon factors}
\begin{figure}
\centering
\includegraphics[trim=4cm 6.5cm 0cm 1cm, clip=true,width=12cm]{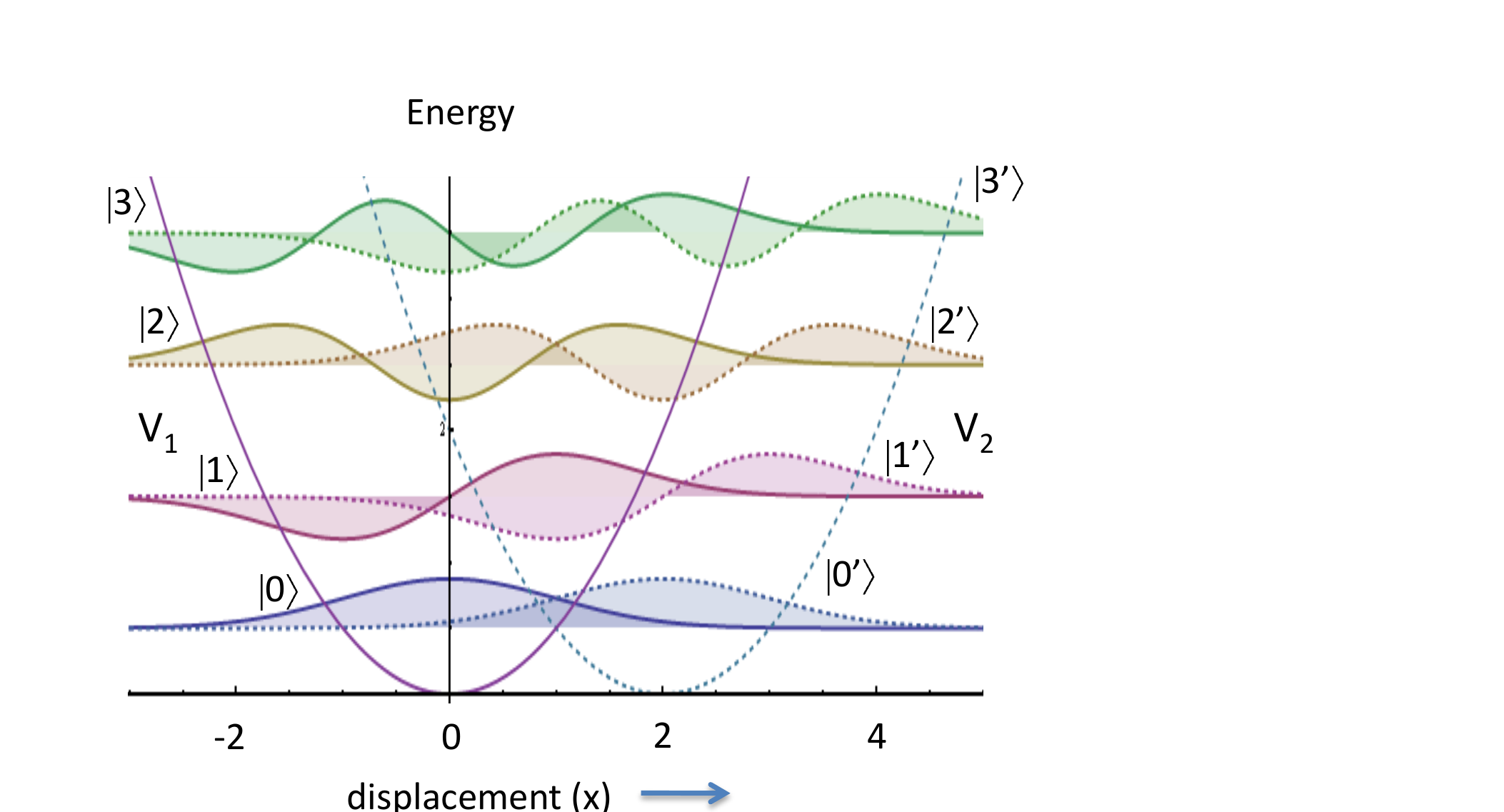}
\caption{
A harmonic potential centred at the origin ($V_1$), a displaced harmonic potential centred at $x=2$ and with $\Delta E = 0$ ($V_2$), and their corresponding wavefunctions.  The unshifted eigenfunctions are
labelled in the computational basis and the shifted eigenfunctions are indicated with primes.
} 
\label{hp}
\end{figure}
In most of the vibronic transitions a molecule undergoes
a transition from one electronic state to another, both of which 
are described by smooth potential energy surfaces consisting of vibrational levels \cite{demtroder2006atoms}.
For low-lying vibrational levels, the potential energy surfaces along each vibrational
degree of freedom can be approximated by simple harmonic potentials.
Considering one vibrational degree of freedom, we model
the electronic ground and excited vibrational levels as eigenstates of two harmonic potentials,
$V_1$ and $V_2$ respectively (Figure \ref{hp}).  For simplicity, we choose the potentials $V_1 = x^2/2$ and $V_2 = (x-b)^2/2 + \Delta E$, 
which are identical up to an overall displacement $b$ in position and/or in energy $\Delta E$. 
Further, we have used unit mass and unit angular frequency, and therefore the kinetic
energies $p^2/2$ will also be identical.  
Denoting ${\mathbf a} = ({\mathbf x}+i{\mathbf p})/\sqrt{2}$ and $\textbf{a}^\dagger = (\textbf{x}-i\textbf{p})/\sqrt{2}$ 
respectively as the annihilation and creation operators, we choose
the eigenstates $\{\vert n \rangle\}$ of the number operator $\textbf{N} = \textbf{a}^\dagger \textbf{a}$
as our computational basis \cite{shankar1994principles}.  
Thus the vibrational Hamiltonians for the two electronic states are
\begin{eqnarray}
{\cal H}_1 &=& \textbf{p}^2/2 + \textbf{x}^2/2 \nonumber \\
&=& \textbf{a}^\dagger \textbf{a} + 1/2  ~~~~ \mathrm{and,} \nonumber \\
{\cal H}_2 &=& \textbf{p}^2/2 + (\textbf{x}-b)^2/2 +  \Delta E \nonumber \\
&=& (\textbf{p}^2/2 + \textbf{x}^2/2) + b^2/2 - \textbf{x}b +  \Delta E \nonumber \\
&=& \textbf{a}^\dagger \textbf{a} + 1/2 + b^2/2 - (\textbf{a}+\textbf{a}^\dagger)b/\sqrt{2} +  \Delta E,
\end{eqnarray}
where the reduced Planck's constant $\hbar$ is set to unity.
A general vibronic excitation constitutes a transition from
a vibrational level $\vert m \rangle$ of the electronic ground state to a vibrational level $\vert n' \rangle$
of the electronic excited state.  
For each vibronic transition $\ket{m} \rightarrow \ket{n'}$,
Franck-Condon principle assigns
the probability,   
\begin{eqnarray}
W_{m,n'} \propto f_{m,n'}(b),
\label{w}
\end{eqnarray}
where
\begin{eqnarray}
f_{m,n'}(b) &=& \vert \langle m \vert n' \rangle \vert^2 \nonumber \\
&=&
\begin{array}{|c|}
\int\limits_{-\infty}^{\infty} \psi_m^*(x) \psi_{n'}(x,b) dx  
\end{array}^2,
\end{eqnarray}
are known as FCFs \cite{demtroder2006atoms}.  
Here $\psi_m(x)$ and $\psi_{n'}(x,b)$ are the position wave-functions
corresponding to the states $\ket{m}$ and $\ket{n'}$ respectively.
Since we have chosen identical potentials with unit angular
frequencies and unit masses, FCFs depend only on 
the displacement $b$.  The separation in energy $\Delta E$ 
only shifts the eigenvalue corresponding to $\ket{n'}$ and
does not affect the FCFs, and hence $\Delta E$ is set to zero.
Theoretical FCFs corresponding to lowest four levels 
of harmonic potentials $V_1$ and $V_2$ are tabulated below. \\

\begin{table}
$\begin{array}{l|l}
\hline
f_{0,0'} & e^{-b^2/2} \\
f_{0,1'} & e^{-b^2/2} ~ b^2/2  \\
f_{0,2'} & e^{-b^2/2} ~ b^4/8 \\
f_{0,3'} & e^{-b^2/2} ~ b^6/48 \\
f_{1,1'} & e^{-b^2/2} ~ (b^2-2)^2/4 \\
f_{1,2'} & e^{-b^2/2} ~ (b^3-4b)^2/16\\
f_{1,3'} & e^{-b^2/2} ~ (b^4-6b^2)^2/96\\
f_{2,2'} & e^{-b^2/2} ~ (b^4-8b^2+8)^2/64\\
f_{2,3'} & e^{-b^2/2} ~ (b^5-12b^3+24b)^2/384\\
f_{3,3'} & e^{-b^2/2} ~ (b^6-18b^4+72b^2-48)^2/2304\\
\hline
\end{array} $
\caption{Analytical forms of FCFs as a function of displacement $b$ 
for a pair of infinite-level identical harmonic oscillators.}
\label{tab1}
\end{table}

\subsection{Estimation of FCFs}
Estimation of FCF, $f_{m,n'}$, is equivalent to measurement of expectation
value of the projection $P_m = \vert m \rangle \langle m \vert$
after preparing the system in excited state $\vert n' \rangle$ since,
\begin{eqnarray}
f_{m,n'} &=& \langle n' \vert m \rangle \langle m \vert n' \rangle \nonumber \\
&=& \langle P_m \rangle_{n'}.
\label{fcexp}
\end{eqnarray}
FCFs can also be obtained, as described in the next section, by extracting the diagonal elements 
of the density matrix after preparing the state $\outpr{n'}{n'}$.

Quantum simulation of a single harmonic potential has earlier
been carried out by Cory and co-workers \cite{somaroo1999quantum}.
Here the vibrational levels of the electronic ground
state are encoded onto the spin states such that
$\ket{0} = \ket{000}$, $\ket{1} = \ket{001}$, $\ket{2} = \ket{010}$, 
$\ket{3} = \ket{011}$, and so on.
Thus the preparation of vibrational level $\vert n \rangle$ of the electronic 
ground state is equivalent to preparing a specific spin state 
and can be achieved by standard methods as explained in the next section.  

The preparation of excited state $\vert n' \rangle$ can be achieved by 
first initializing the system
in the corresponding  state $\vert n \rangle$ of the electronic
ground state and translating it in position from origin ($x=0$)
to the point $x=b$.  
This translation
can be achieved by the unitary operator
\begin{eqnarray}
U_T(b) = e^{-i \textbf{p} b}.
\end{eqnarray}
Discrete translations up to $b_0 \ge 0$ can be achieved by repetitively
applying the operator $U_T(b_0/N)$ such that
$U_T(b_0 k /N) = [U_T(b_0/N)]^k$, where $k \le N$ are non-negative integers.
In the following we describe two experimental techniques to measure FCFs, 
one based on direct measurement of projection operators using diagonal density matrix
tomography and the other based on Moussa protocol.

\begin{figure}[b]
\centering
\includegraphics[trim=6.5cm 5cm 5cm 4cm, clip=true,width=5cm]{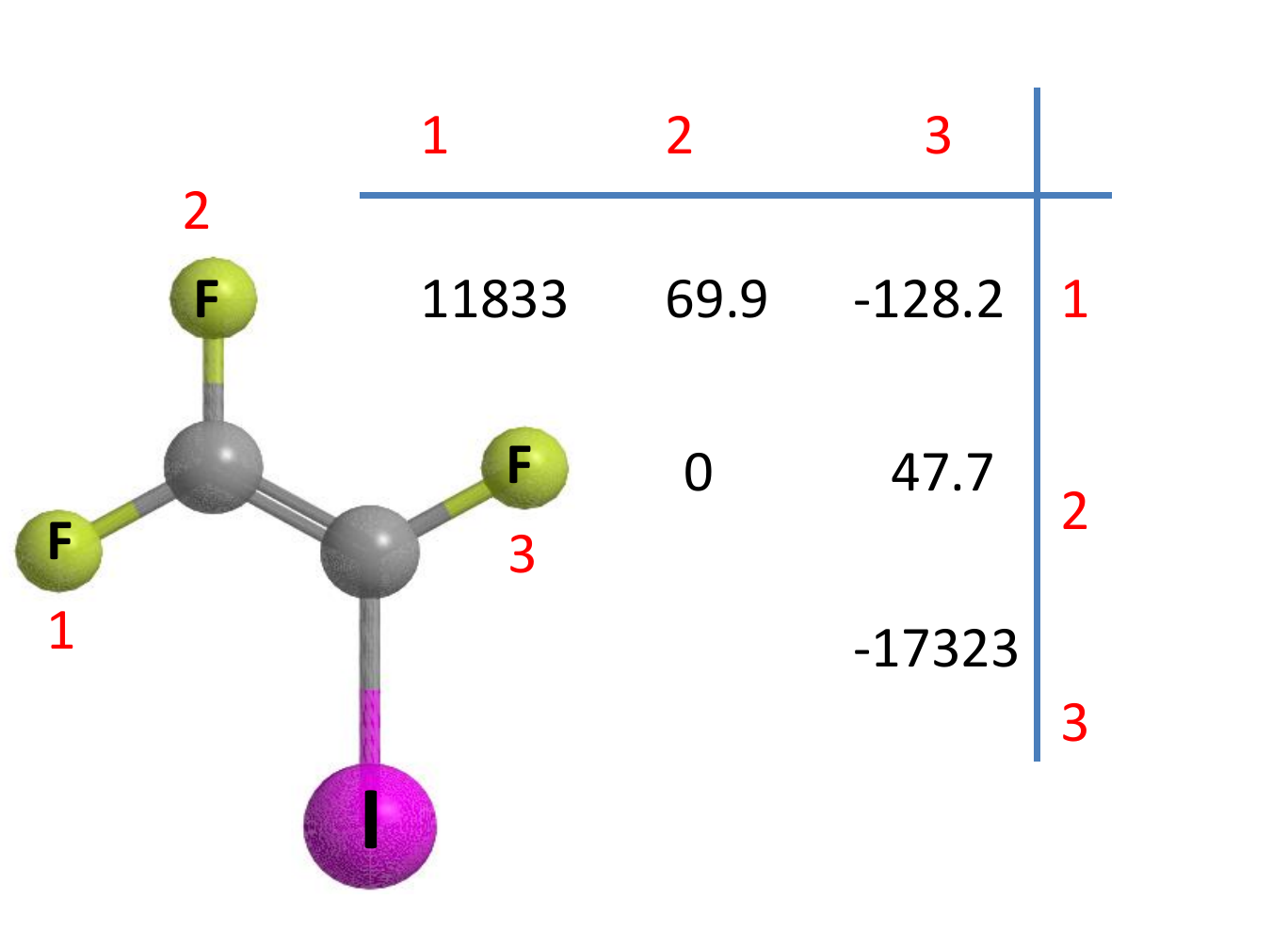}
\caption{
Molecular structure of iodotrifluoroethylene and its Hamiltonian 
parameters.  The diagonal elements represent relative resonance  
frequencies and the off-diagonal elements represent the strengths
of scalar (J) couplings.
} \label{molecule}
\end{figure}

\section{Experiments}
We have chosen the three spin-1/2 $^{19}$F nuclei of iodotrifluoroethylene
(C$_2$F$_3$I) dissolved in acetone-D$_6$ as our three-qubit NMR quantum processor.
All the experiments are carried out on a 500 MHz Bruker NMR spectrometer 
at an ambient temperature of 300 K.  The molecular structure and the 
Hamiltonian parameters of this system are given in Figure \ref{molecule}.
In the following we describe two experimental methods for estimating FCFs.

\subsection{Using diagonal density matrix tomography}
The three-spin system  C$_2$F$_3$I provides us with a 3-qubit
quantum register which can be used to encode the lowest eight levels
of a harmonic oscillator.  To determine all FCFs corresponding
to the lowest four levels, 
we carry out a set of four experiments each with a different
initial state
$\rho_\mathrm{ini} = \outpr{n}{n} \in \{\ket{000},\ket{001},\ket{010},\ket{011}\}$ 
encoding a particular vibrational level of the electronic
ground state.  As is customary in NMR quantum information studies,
each pure state is mimicked by preparing a corresponding pseudopure state which
in turn is prepared by standard methods \cite{cory,sup}.

As explained earlier, FCF $f_{m,n'}(b_0 k/N) = \rho'_{m} = \mprod{m}{\rho'}{m}$
is a diagonal element of the translated state 
$\rho' = U_T(b_0 k/N) \outpr{n}{n} U_T^\dagger(b_0 k/N)$.
In our experiments we have used $b_0 = 3$, $N = 11$, and integer $k\in[0,11]$.
The complete translation operator was realized using amplitude and 
phase modulated radio frequency pulses.  
These pulses were designed via GRadient Ascent Pulse
Engineering  (GRAPE) technique \cite{khaneja2005optimal} and had 
average Hilbert-Schmidt fidelities above $0.99$ over a
spatial RF inhomogeneity in the range 90\% to 110\% of the nominal field.

The diagonal density matrix tomography involves destroying all the 
off-diagonal elements of $\rho'$ using a pulsed-field-gradient (PFG)
followed by linear detection using a small flip-angle (6 degree) y-pulse.
The detection pulse mixes the diagonal elements $\rho'_m$ 
into observable coherences which appear as various resolved 
transitions in the NMR signal.
The real part $r_j$ of $j$th transition normalized w.r.t. the
reference (corresponding to equilibrium density matrix followed by 
the linear detection) constitutes a linear combination 
$r_j = \sum_m M_{j,m} \rho'_m$, where the coefficients 
$M_{j,m}$ form a $13\times8$ dimensional constraint matrix \cite{sup}.
The diagonal elements $\rho'_{m}$ can be extracted by solving
the overdetermined set of linear equations \cite{sup}.

\begin{figure}
\centering
\hspace*{-.3cm}
\includegraphics[trim=1.5cm .2cm 1.5cm 0.1cm, clip=true,width=9cm]{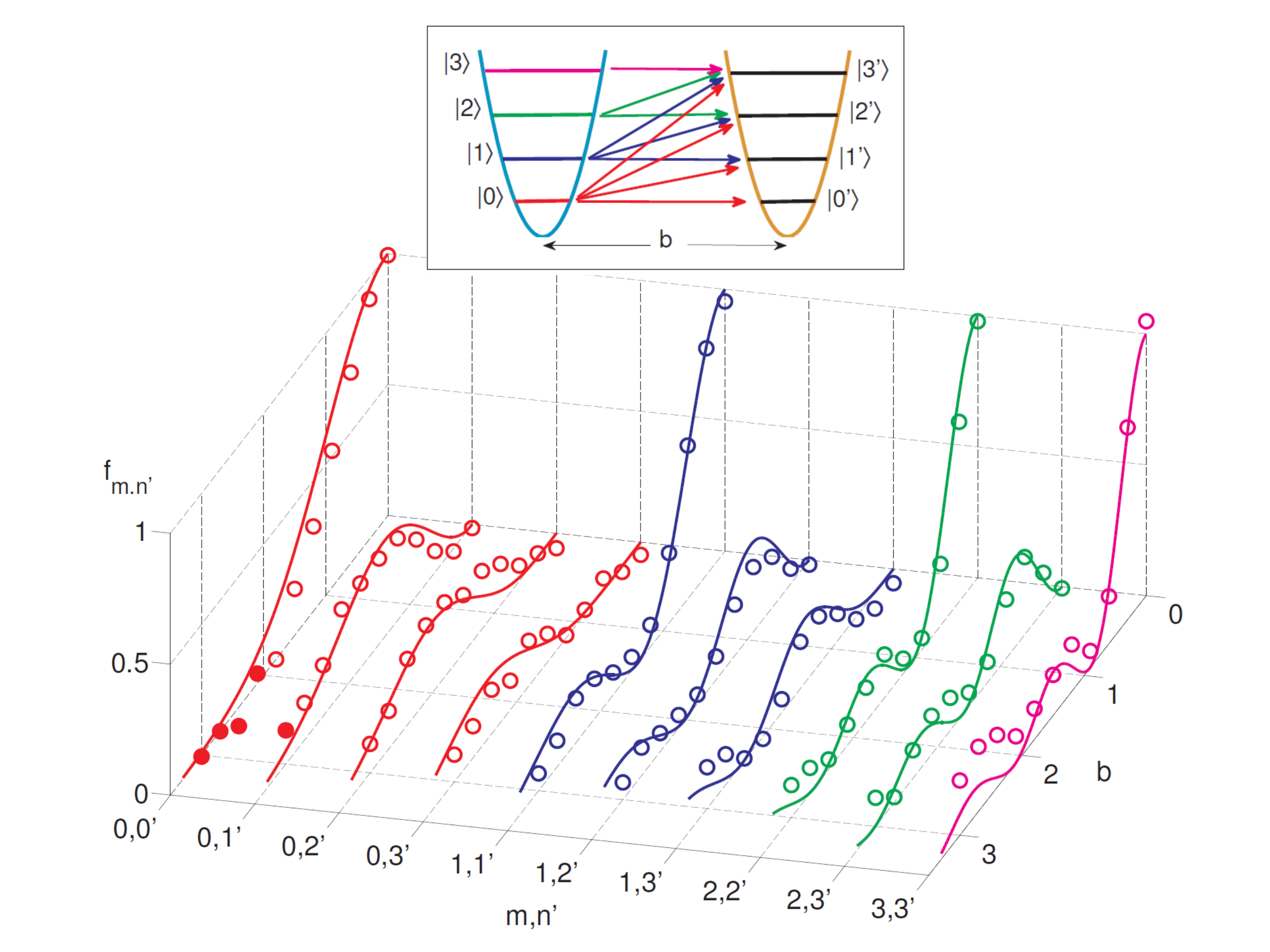}
\caption{
Experimental FCFs (circles) corresponding to lowest four-levels of Harmonic oscillators (shown in inset)
measured using diagonal density matrix tomography of 3-qubits.  
Analytical forms of FCFs (solid-line) for infinite-level potentials (Table \ref{tab1}) are also plotted for comparison.
are shown for comparison. All the FCFs are plotted versus displacement $b$. Filled circles indicate the FCFs 
which are entirely due to the overlap of tunnelling amplitudes.
} 
\label{fcall}
\end{figure}

The experimental values of all FCFs as a function of $b$ for the lowest four levels of
a pair of Harmonic oscillator  are shown by circles in Fig. \ref{fcall}.
The theoretical curves corresponding to infinite-level harmonic
oscillators, are also shown for comparison.  
  Since the FCFs are symmetric w.r.t. the exchange
of subscripts, it can be noted that $f_{m,n'} = f_{n,m'}$.
While there is a good
agreement between the theoretical and experimental data, the errors
are mainly due to imperfections in preparation of pseudopure states, imperfections
in implementing the translation operators, and decoherence.
The filled 
circles in Fig. \ref{fcall} correspond to the overlap between the  
wavefunctions beyond the classical turning points, i.e., $b \ge 2$ for $f_{0,0'}$
and $b \ge 1+\sqrt{3}$ for $f_{0,1'}$.  
Thus the non-zero values of FCFs at these points provide a direct way of experimental 
observation of quantum tunnelling.

\subsection{Using Moussa protocol}
\begin{figure}[b]
\centering
\includegraphics[trim=4.8cm 9cm 0cm 3cm, clip=true,width=10cm]{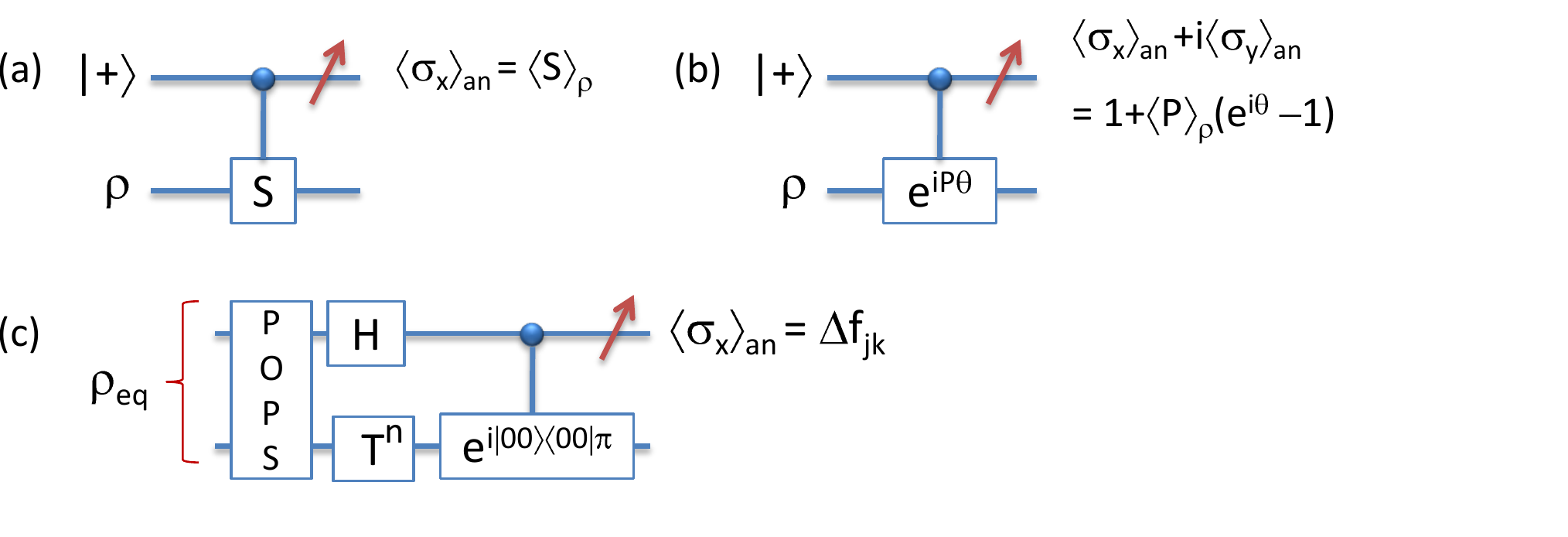}
\caption{Circuits for Moussa protocol (a), its extension (b), and
the overall circuit for estimating FCFs.
In all the cases the top qubit corresponds to ancilla and the remaining
are system qubits.
In (b) $P$ is a projection observable. 
} 
\label{moussa}
\end{figure}

As described in eqn. \ref{fcexp}, estimating FCFs is equivalent to measurement of
certain projections.
An existing method, known as Moussa protocol \cite{moussa2010testing}, achieves the measurement of 
any unitary observable with the help of an ancilla qubit. 
The circuit implementation of Moussa protocol is illustrated in Figure \ref{moussa}a.
It involves an ancilla qubit initialized in state $\vert + \rangle$ and
the system is initialized in any desired state $\rho$.  In order to
measure the expectation value of a unitary observable $S$, we need
to apply the operator on the system controlled by ancilla as shown 
in Figure \ref{moussa}a.  Then we obtain $\langle S \rangle$ by directly measuring
$\sigma_x$ operator on the ancilla \cite{moussa2010testing}, i.e.,
\begin{eqnarray}
\langle S \rangle_\rho = \langle \sigma_x \rangle_\mathrm{an}.
\end{eqnarray}

Here we extend this protocol to extract expectation values of a projection operator
which can then be used to measure any general observable.

Let $P$ be a projection operator,  with $P^n = P$, for any
positive integer $n$. Consider a
unitary operator $S = \exp(i P \theta)$ with a chosen
angle $\theta$.  Consider the circuit shown in Figure 
\ref{moussa}b.
It can be shown that the expectations of ancilla are 
related to the $S$ by \cite{sup}
\begin{eqnarray}
\expv{\sigma_x}_\mathrm{an} = \expv{S+S^\dagger}_\rho/2, ~~ \mathrm{and} ~~
\expv{i\sigma_y}_\mathrm{an} = {S-S^\dagger}_\rho/2, \nonumber
\end{eqnarray}
so that
\begin{eqnarray}
\expv{\sigma_x}_\mathrm{an}
+
i\expv{\sigma_y}_\mathrm{an}
= \langle S \rangle_\rho 
= \langle e^{i  P  \theta} \rangle_\rho 
= 1 + \langle P \rangle_\rho (e^{i\theta}-1), \nonumber
\end{eqnarray}
and hence
\begin{eqnarray}
\langle P \rangle_\rho = \frac{1-\langle \sigma_x \rangle_\mathrm{an}}{1-\cos{\theta}}.
\end{eqnarray}
Taking $\theta = \pi$, we obtain $\langle P \rangle_\rho = (1-\langle \sigma_x \rangle_\mathrm{an})/2$.

In our experiments, we have selected $F_1$ qubit (see Figure \ref{molecule}) as the
ancilla and the other two qubits for representing the lowest four levels
of the Harmonic oscillator.  As explained in the previous section, we
first prepare the system in one of the four levels of the electronic
ground state and translate it in position towards the corresponding 
excited state.  Instead of initializing the system into a single
level, we utilize the `Pair Of Pseudopure States' (POPS) method \cite{fung2001use}
which prepares the system into a traceless pseudopure mixture of two levels
$\rho_{j,k} = \vert j \rangle \langle j \vert - \vert k \rangle \langle k \vert$, 
where $j,k(\neq j) \in \{00,01,10,11\}$ \cite{sup}.
After applying the translation operator, we obtain the corresponding
excited states $\rho_{j,k}' = \vert j' \rangle \langle j' \vert - \vert k' \rangle \langle k' \vert$. In our experiments, we have selected the
projection operator $P_{00} = \outpr{00}{00}$ as our observable.  This way,
we determine the overlap between the level $\ket{00}$ of the electronic
ground state with $\ket{n'(b)}$ of the electronic excited state, and 
obtain the FCFs $f_{00,k'}$.

The complete circuit with initialization and extended Moussa protocol is shown 
in Figure \ref{moussa}c.  
Here the first qubit is the ancilla and the other two qubits form the system.  
We first applied the translation operator $U_T(b_0 k/N)$ after preparing the
initial state $\outpr{+}{+}\otimes \rho_{jk}$.  
In these experiments $b_0$ was set to $4$ and $N$ to $11$.  
Again, the complete translation operator was realized using amplitude and 
phase modulated radio frequency pulses having fidelities over 0.995.  
Then the operator $\exp(i P_{00} \pi)$ on the system qubits controlled by the
ancilla qubit was applied. The GRAPE pulse of this operator had
an average fidelity of 0.989 over the similar RF inhomogeneity range.
Finally the real part of the ancilla signal 
that is proportional to $\expv{\sigma_x}_\mathrm{an}$ is measured
by integrating the ancilla transitions.  Thus after each 
preparation $\rho_{j,k}'$ we can extract the difference of FCFs 
$\Delta f_{j,k} = f_{00,j'}-f_{00,k'}$.  In order to determine all
individual FCFs of the form $f_{00,k'}$, we perform four different experiments and measure
$\Delta f_{00,01'}$, $\Delta f_{00,10'}$, $\Delta f_{01,11'}$, and
$\Delta f_{10,11'}$, and solve the four linear equations with the help of 
an additional constraint given by the normalization condition $\sum_j f_{00,j'} = F$.

\begin{figure}
\centering
\includegraphics[trim=1.5cm .3cm 0cm 1cm, clip=true,width=8.7cm]{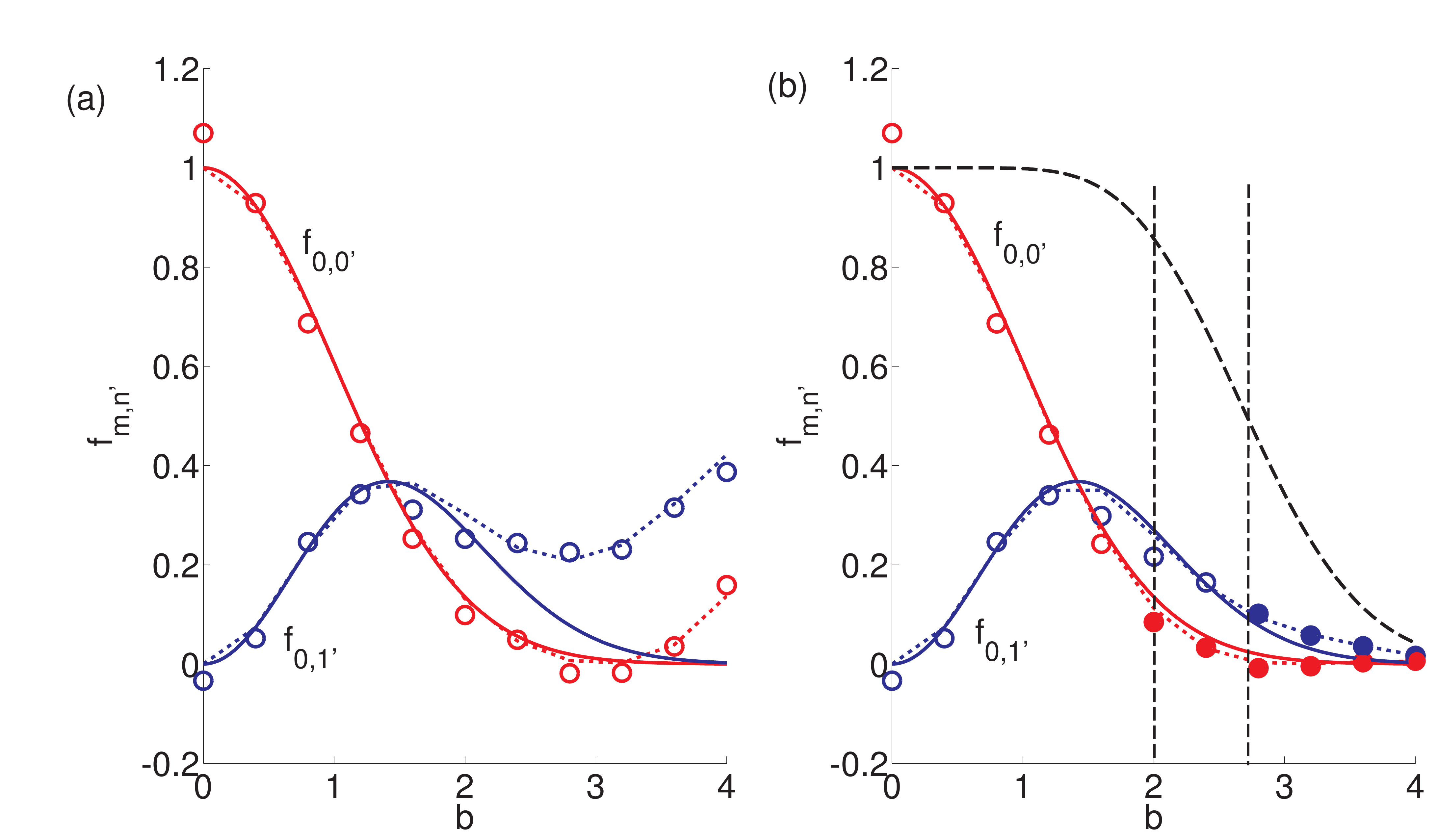}
\caption{
Experimental FCFs (circles) correspoding to 4-level harmonic oscillators (encoded by two qubits) 
versus the displacement $b$ obtained using Moussa protocol.
The simulated FCFs (dotted lines) for the 4-level system and
analytical FCFs (smooth lines) for infinite-level system (Table \ref{tab1}) are also shown 
for comparison.
The dashed curve at the top of (b) corresponds to the
normalization used.  The thin vertical dashed lines at $b=2,~1+\sqrt{3}$
mark the beginning of classically forbidden regions for $f_{0,0'},
f_{0,1'}$ respectively.
} 
\label{results}
\end{figure}

The results of the experiment with $F=1$ and for various values of $b\in[0,4]$ 
are shown in Figures \ref{results}a.
The experimental data points are shown by symbols.  To understand the
systematic errors in the experimental data, we have simulated FCF values
using lowest four levels of the computational basis.  
In Figures \ref{results} The simulated curves 
are shown by dotted lines and the expected FCF curves for infinite level systems
are shown by smooth lines.
Here a good agreement between the experimental results and the
simulated curves can be observed. However,
as can be readily seen, the observed FCFs
gradually deviate away from the infinite case with increasing values of $b$.
The deviation between the finite and infinite FCFs is due to the effect of
truncation in the basis-set, and can be minimized by increasing the number
of system qubits \cite{sup}.

If one has the prior information on the normalization condition $F$,
it is possible to get better agreement with the infinite case.  This is
illustrated in Figures \ref{results}b.
The total probability for the first four levels in the infinite-level case varies with $b$ as
$\sum_j f_{00,j} = \left[1+b^2/2+b^4/8+b^6/48\right]e^{-b^2/2}$ 
and is shown by thin-dashed lines at the top of Figure \ref{results}b.
The experimental results with this normalization are shown by symbols.
With this normalization, the FCFs show
considerably good agreement with the infinite-level case.
Again FCFs in classically forbidden regions are shown by filled circles
in \ref{results}b.

\section{Conclusions}
We formulated a general procedure for estimating Franck-Condon Factors (FCFs) 
in a quantum information processor.  
First we noted that an excited state can be prepared by spatially translating
the corresponding ground state. Secondly, we observed that estimation
of FCFs is equivalent to the measurement of certain projection operators
on a set of qubits encoding the excited state of the system. 
We described two techniques of extracting the expectation values: one based
on diagonal tomography and the other using an extended Moussa protocol.  
We illustrated both of these techniques using a three qubit NMR quantum simulator
and compared the results with the simulated curves for finite 
and infinite level systems.  We demonstrated that even with only three/two qubits
encoding the system, the FCFs corresponding to the lowest four/two levels
still matched fairly well with the infinite-level theory.
In general, if we have an arbitrary potential for the excited state,
then it may still be possible to prepare the final state using adiabatic 
evolutions.

\bibliographystyle{apsrev4-1}
\bibliography{fc1}

\section*{Supplementary material}
\subsection{Finding expectation value of a general unitary operator using Moussa protocol}
Consider a general unitary operator $U$.  To determine its expectation value 
we prepare the system in any desired state $\rho$, ancilla in state $\vert + \rangle$,
and apply the Moussa protocol as shown in Fig. 2.  The state of the combined system
evolves under the controlled-$U$ operation as
\begin{eqnarray}
\outpr{+}{+} \otimes \rho \stackrel{U}{\longrightarrow} 
\frac{1}{2} 
\{ 
\outpr{0}{0} \otimes \rho +
\outpr{0}{1} \otimes \rho U^\dagger + \nonumber \\
\outpr{1}{0} \otimes U \rho +
\outpr{1}{1} \otimes U \rho U^\dagger
\}.
\end{eqnarray}
After tracing out system, the ancilla becomes
\begin{eqnarray}
\frac{1}{2}
\left\{ 
\outpr{0}{0} +
\outpr{0}{1} \expv{U^\dagger} +
\outpr{1}{0} \expv{U} +
\outpr{1}{1} 
\right\} = \nonumber \\
\left[
\begin{array}{c c}
1 & \expv{U^\dagger} \\
\expv{U} & 1
\end{array}
\right].
\end{eqnarray}
In an NMR signal obtained with a quadrature detection, the real and imaginary parts 
correspond to the expectation values of $\sigma_x$ and $\sigma_y$ observables
respectively [Ref].
Thus the real and imaginary parts of the ancilla signal are proportional to 
$\expv{\sigma_x} = \expv{\frac{U+U^\dagger}{2}}$ and
$\expv{i\sigma_y} = \expv{\frac{U-U^\dagger}{2}}$, from which the expectation value 
$\expv{U} = \expv{\sigma_x} + i\expv{\sigma_y}$ can be extracted.

\subsection{Extracting expectation value of a Hermitian operator compatible with a pure-state}
The circuit for this measurement is shown in Figure 3.
Given the Hermitian operator $A$ to be compatible with a pure state $\rho$, we can
simultaneously diagonalize both of these using a similarity transformation $U_s$,
i.e., $A_d = U_s^\dagger A U_s$, $\rho_d = U_s^\dagger \rho U_s$ are diagonal, and
therefore $\expv{A_d}_{\rho_d} = \expv{A}_{\rho}$.  
As shown above, the complex ancilla signal corresponds to
$\expv{\sigma_x}_\mathrm{an}+i\expv{\sigma_y}_\mathrm{an}
=\expv{e^{i A_d \theta}}_{\rho_d}$.  In the following we prove that 
$\expv{e^{i A_d \theta}}_{\rho_d} = e^{i \expv{A_d}_{\rho_d} \theta}$.
Since the system is in a pure state, we can write $\rho_d = \outpr{n}{n}$ and
$(\rho_d)_{ij} = \delta_{in}\delta_{jn}$.  
Now consider,
\begin{eqnarray}
\expv{A_d^k}_{\rho_d} &=& \mathrm{Tr}\left[
\rho_d A_d^k
\right] 
\nonumber \\ &=&
\sum_{i}(\rho_d A_d^k)_{ii}
\nonumber \\ &=&
\sum_{ij_{1}j_{2}\cdots j_{k}}\delta_{in}\delta_{j_1 n} 
(A_d)_{j_1j_2} \cdots (A_d)_{j_{k} i}
\nonumber \\ &=&
(A_d)_{nn}^k, ~~~ \mathrm{since} ~ A_d ~ \mathrm{is ~ diagonal.}
\end{eqnarray}
For $k = 1$, we have $\expv{A_d}_{\rho_d} = (A_d)_{nn}$, and therefore
$\expv{A_d^k}_{\rho_d} = \expv{A_d}_{\rho_d}^k$, which implies, 
$\expv{e^{i A_d \theta}}_{\rho_d} = e^{i \expv{A_d}_{\rho_d} \theta}$.
Since $\expv{A_d}_{\rho_d} = \expv{A}_{\rho}$, we have
$\expv{\sigma_x}_\mathrm{an}+i\expv{\sigma_y}_\mathrm{an}
=e^{i \expv{A}_\rho \theta}$.

\begin{figure}[h]
\centering
\includegraphics[trim=3cm 4cm 0cm 4cm, clip=true,width=6cm]{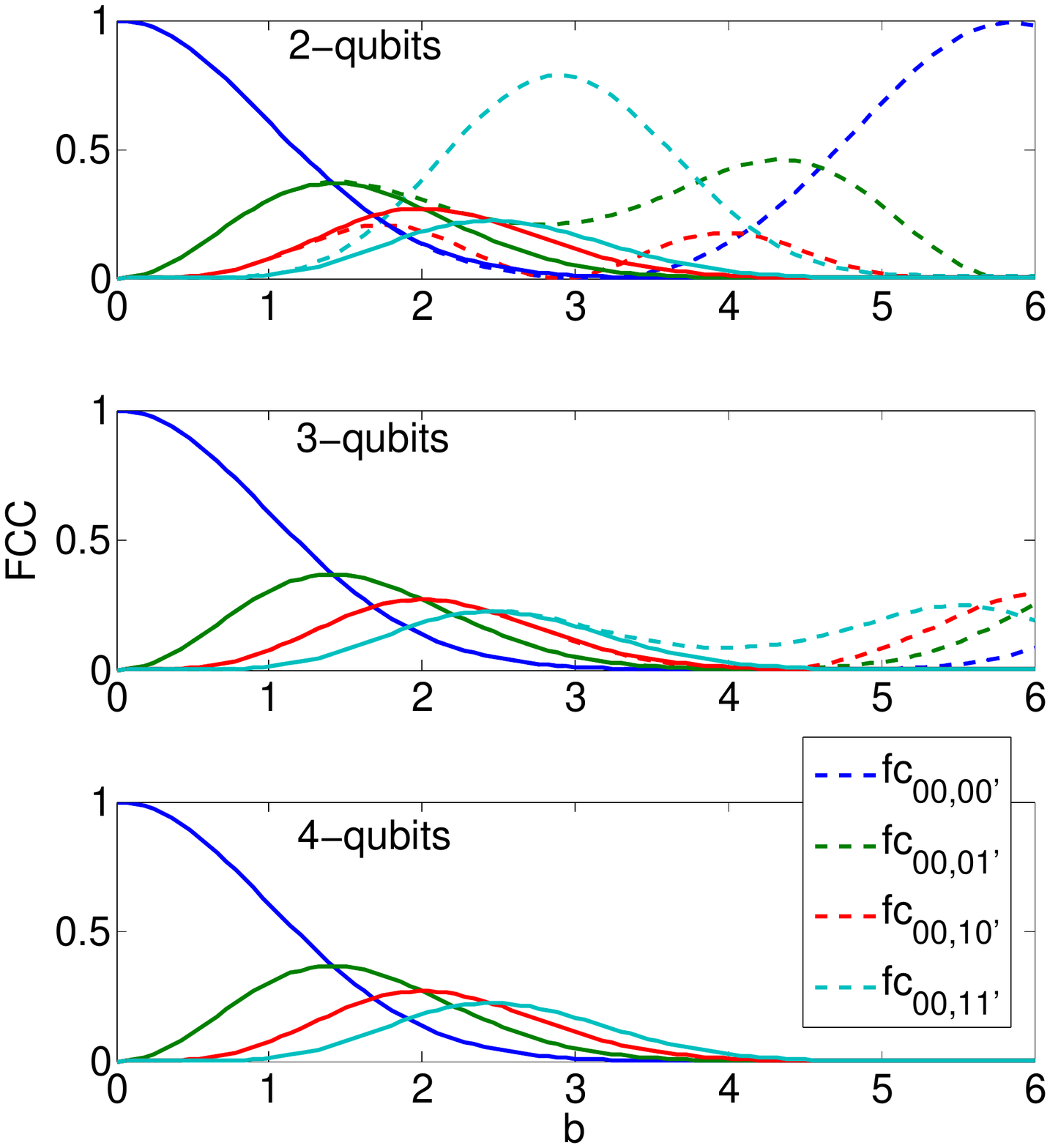}
\caption{Numerical simulation of FCFs to show the effect of truncation of
basis (i.e., finite levels of the potentials).
4-basis states (top-panel), 8-basis states (middle-panel), and 16-basis states (bottom-panel)
are used to simulate FCFs.
} \label{ref}
\end{figure}

\textit{Hermitian operator compatible with a pure-state:}
Let the system be in a pure state $\rho$ and $A$ be a Hermitian
operator compatible with $\rho$, i.e., they have a common
eigenbasis.  We can transform $\rho$ into its eigenbasis 
using a similarity transformation $U_s$, such that 
$\rho = U_s \rho_d U_s^\dagger$, $A = U_s A_d U_s^\dagger$,
where $\rho_d$ and $A_d$ are diagonal.
Again we can perform the Moussa protocol (Figure \ref{moussa}c) by applying the
unitary $S = \exp(i A_d \theta)$ controlled by ancilla, and we
obtain \cite{sup},
\begin{eqnarray}
\langle \sigma_x \rangle_\mathrm{an} 
+i\langle \sigma_y \rangle_\mathrm{an}
= e^{i  \langle A\rangle_{\rho}  \theta}.
\end{eqnarray}
The expectation $\langle A\rangle_{\rho}$ can be extracted
by knowing the rotation angle $\theta$.

Numerical simulations on the effect of finite-basis is illustrated in Fig. \ref{ref}.

\begin{figure}[h]
\centering
\includegraphics[trim=0cm 0cm 0cm 0cm, clip=true,width=6cm]{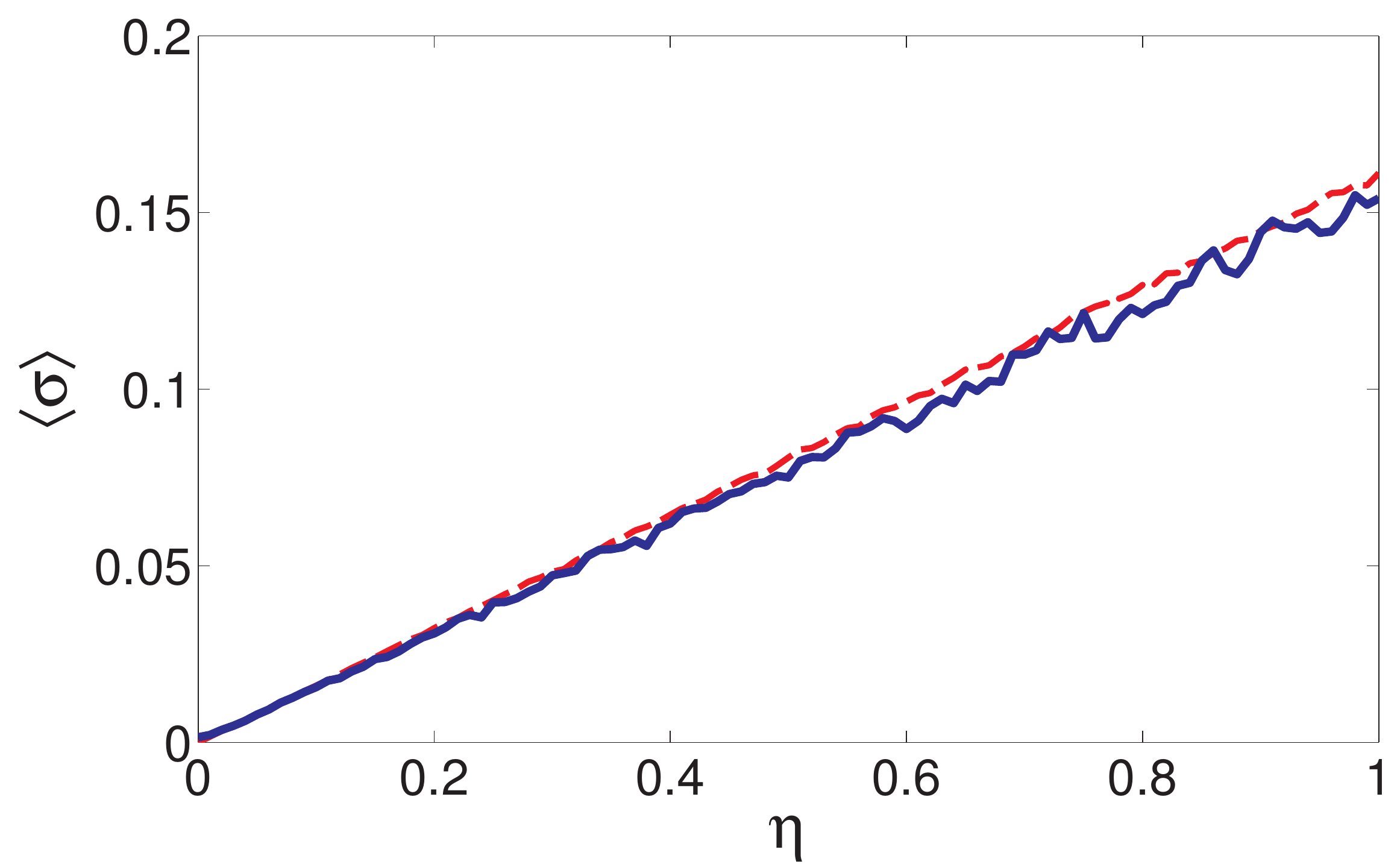}
\caption{
Average standard deviation of simulated FCFs as a function of noise amplitude $\eta$.
The blue solid-line corresponds to the diagonal-tomography method and the red dashed-line
corresponds to Moussa protocol.
} 
\label{robust}
\end{figure}

Finally to demonstrate the robustness of these methods, calculated
FCFs by adding random numbers (between $-\eta$ and $\eta$) to simulated the NMR intensities (see Fig. \ref{robust}).  The average standard
deviation $\ket{\sigma_\epsilon}$ of FCFs as a function of noise amplitude $\eta$ is shown in Figure \ref{robust}.
It can be observed that $\ket{\sigma_\epsilon} < 0.2$ for both the methods even at $\eta=1$, indicating the robustness of 
the methods.

\end{document}